\documentclass[a4paper]{article}

\usepackage[utf8]{inputenc} 
\usepackage{hyperref}       
\usepackage{url}            
\usepackage{amsfonts}       
\usepackage{amsmath}
\usepackage{graphicx}
\usepackage{caption}
\usepackage{float}
\usepackage[left=3cm,right=3cm,top=2cm,bottom=2cm]{geometry}
\usepackage[english]{babel}
\usepackage{amsmath}
\usepackage{graphicx}
\usepackage[colorinlistoftodos]{todonotes}
\usepackage{float}
\usepackage{dsfont}
\usepackage{bm} 
\usepackage{url}
\usepackage{booktabs} 
\usepackage{multicol}
\usepackage{multirow}
\usepackage{gensymb}
\usepackage{tabularx}
\usepackage{cite}
\DeclareCaptionFormat{sanslabel}{#3}%

\begin{document}

\Large
\begin{center}
\vspace{1cm}
\textbf{Dynamic full-field optical coherence tomography: 3D live-imaging of retinal organoids}

\vspace{0.5cm}

\normalsize
Jules Scholler$^{1,\dag}$, Kassandra Groux$^{1,\dag}$, Olivier Goureau$^2$, José-Alain Sahel$^{2,3,4,5}$,  Mathias Fink$^1$, Sacha Reichman$^2$, Claude Boccara$^1$ and Kate Grieve$^{2,3,*}$

\end{center}
\scriptsize
$^1$Institut Langevin, CNRS, ESPCI Paris, PSL Research
University, 10 rue Vauquelin, Paris, France\\
$^2$Institut de la Vision, Sorbonne Université, INSERM, CNRS, F-75012, Paris, France\\
$^3$Quinze-Vingts National Eye Hospital, 28 Rue de Charenton, Paris, 75012, France\\
$^4$Fondation Ophtalmologique Rothschild, F-75019 Paris, France \\
$^5$Department of Ophthalmology, The University of Pittsburgh School of Medicine, Pittsburgh, PA 15213, United States \\
$^\dag$These authors contributed equally to this work \\
$^*$kategrieve@gmail.com

\normalsize


\begin{abstract} 
Optical coherence tomography offers astounding opportunities to image the complex structure of living tissue, but lacks functional information. We present dynamic full-field optical coherence tomography to image living human induced pluripotent stem cell-derived retinal organoids non-invasively. Colored images with an endogenous contrast linked to organelle motility are generated, with sub-micrometer spatial resolution and millisecond temporal resolution, opening an avenue to identify specific cell types in living tissue via their function.
\end{abstract}

\vspace{1cm}


The comprehension of the human body and its mechanisms at the sub-cellular scale is still an open area of research. During the seventeenth century, the first examinations of life under the microscope were conducted directly on humans, animals and bacteria \cite{hajdu_first_2002}. Then, at the end of the nineteenth century, cell culture began to replace \textit{in vivo} studies, as this allow creating of \textit{in vitro} models beneficial for the comprehension of biological phenomena in different environments \cite{mazzarello_unifying_1999, jedrzejczak-silicka_history_2017}. Because of the two-dimensional nature of early cell cultures, the possibilities of understanding tissues and organs as a whole was limited. Recently, three-dimensional (3D) cultures have been developed from stem cells to generate organoids that mimic a variety of tissues and serve as models of human development and disease studies \cite{hynds_concise_2013, lancaster_organogenesis_2014, clevers_modeling_2016,hartogh_concise_2016, rossi_progress_2018, lancaster_disease_2019}. Organoids could also serve as sources of human tissues for transplantation  and as platforms for drug screening \cite{fatehullah_organoids_2016,gagliardi_characterization_2018,Gagliardi_2019,BENMBAREK_2019}. These self-organizing structures develop cellular composition and architecture similar to \textit{in vivo} tissues, thereby replicating biologically relevant intercellular phenomenon \textit{in vitro} \cite{hynds_concise_2013, lancaster_organogenesis_2014, clevers_modeling_2016, hartogh_concise_2016, rossi_progress_2018, lancaster_disease_2019,  slembrouck-brec_reprogramming_2019, reichman_generation_2017,Reichman_pnas_2014}.

\begin{figure}
	\centering
    \includegraphics[width=1\linewidth]{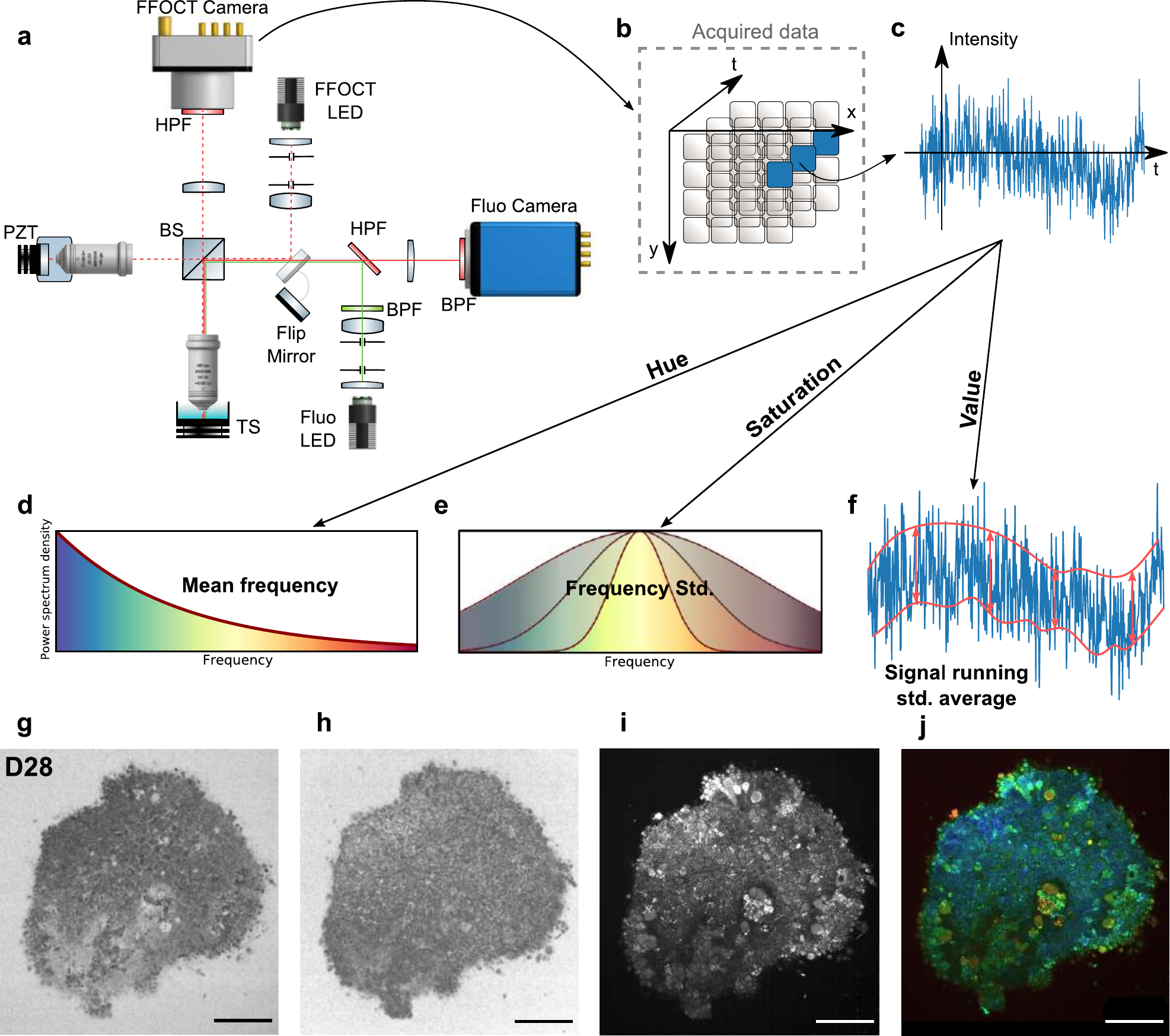}
    \caption{Experimental setup and post-processing schematic. \textbf{Acquisition of images a}, Full-Field OCT setup combined with wide-field fluorescence microscopy (side view). PZT: piezoelectric translation - TS: translation stage - BPF: band pass filter - HPF: high pass filter. As the fluorescence is recorded in the same spectral range as D-FFOCT, the D-FFOCT and fluorescence images could not be recorded at the same time: for this purpose a flip mirror was added to switch from one modality to another. \textbf{b}, 3D cube of data $(x, y, t)$ acquired before processing. Each voxel is processed independently. \textbf{c}, An intensity trace is plotted for a voxel inside a living retinal organoid. \textbf{Post-processing steps d-f}. Dynamic images are computed in the HSV colorspace. \textbf{d}, The hue is computed with the mean frequency, from blue (low temporal frequencies) to red (high temporal frequencies) \textbf{e}, The saturation is computed as the inverse of the frequency bandwidth; as a consequence a signal with a broader bandwidth (e.g. white noise) appears dull whereas a signal with narrow bandwidth will appear vivid. \textbf{f}, The value is computed as the running standard deviation \cite{apelian_dynamic_2016}. Bottom row in a D29 retinal organoid, \textbf{g}, Computation of mean frequency (hue) \textbf{h}, frequency bandwidth (saturation) and \textbf{i}, dynamic (value)  -before \textbf{j}, reconstruction. Scale bar: $50~\mu m$.}
    \label{fig1}
\end{figure}

For each biological trend, optical imaging devices have been developed and optimized to image tissues, cell cultures and recently, organoids, which are one of the most fundamental tools in biology, clinical pathology and medical diagnosis \cite{yuste_fluorescence_2005}. There are many challenges in imaging 3D structures: due to their relatively transparent nature it is hard to obtain contrast on specific structures without staining. Moreover, 3D samples require optical sectioning in order to discriminate the layer in focus from out of focus ones. In this study we present the application of Dynamic Full-Field Optical Coherence Tomography (D-FFOCT) to imaging of retinal organoids derived from human induced pluripotent stem-cells (hiPSC) \cite{reichman_generation_2017} which are a major breakthrough in the study of retina and retinal diseases. These hiPSC-derived retinal organoids are routinely imaged with various techniques (see Supplementary Table 1). However, each of the existing methods present major drawbacks such as the need for fixation or mechanical slicing, hence rendering impossible the study of dynamic phenomena; the need for labelling, requiring cumbersome and costly preparation; or lack of functional contrast, so indicating only cell presence and not cell health or behavior \cite{browne_structural_2017, capowski_reproducibility_2019, reichman_generation_2017, cora_cleared_2019}. Optical Coherence Tomography (OCT) is commonly used in biology and medicine for 3D imaging of micro-structures in tissue. OCT contrast arises from the local endogenous optical backscattering level \cite{huang_optical_1991}. The main drawback of traditional OCT is the trade-off between imaging depth and resolution. In order to increase lateral resolution, the numerical aperture of the system must be increased. As a consequence, the depth of field reduces and only a small layer of the sample can be imaged. Current OCT systems therefore have a lateral resolution on the order of $10~\mu m$, insufficient to resolve cell structures laterally.  Using an incoherent light source and a camera, Full-Field OCT (FFOCT) is an en face variant of OCT with a higher spatial and temporal resolution in the en face plane \cite{dubois_ultrahigh-resolution_2004}. As FFOCT acquires an en face plane rather than a line in the depth direction, the numerical aperture can be arbitrarily increased without any adverse effects on imaging depth. Using the FFOCT experimental setup shown Fig.~\ref{fig1}(a) and detailed in Methods, a novel contrast mechanism has recently been exploited by measuring temporal fluctuations of the backscattered light in a technique called Dynamic FFOCT (D-FFOCT) \cite{apelian_dynamic_2016}. These dynamic measurements reveal sub-cellular structures that are very weak back-scatterers and provide contrast based on local intra-cellular motility \cite{thouvenin_cell_2017, scholler_probing_2019} with sub-micrometer resolution, and can achieve millisecond temporal resolution to study fast phenomena.

In FFOCT, the light coming back from the sample slice of interest interferes with the light coming back from the reference mirror and is projected onto the camera (Fig.~\ref{fig1}(a)). In order to compute a D-FFOCT image, a movie (typically 512 frames) of the interferogram pattern is recorded and processed (see Methods) to extract local fluctuations and render them colored (Fig.~\ref{fig1}(b-j)). Using Hue-Saturation-Value (HSV) colorspace, image brightness is linked to fluctuation amplitude while color is linked to fluctuation speed, from blue (slow) to red (fast) through green (in between). By translating the sample in the axial direction to acquire a stack of planes, see Supplementary vid. 1, a 3D volume can be reconstructed (see Methods). Alternatively, a series of dynamic images may be acquired in the same plane to follow the evolution of activity over several hours in a time-lapse fashion with a temporal resolution of up to $20~ms$ (see Methods).

\begin{figure}
	\centering
    \includegraphics[width=0.8\linewidth]{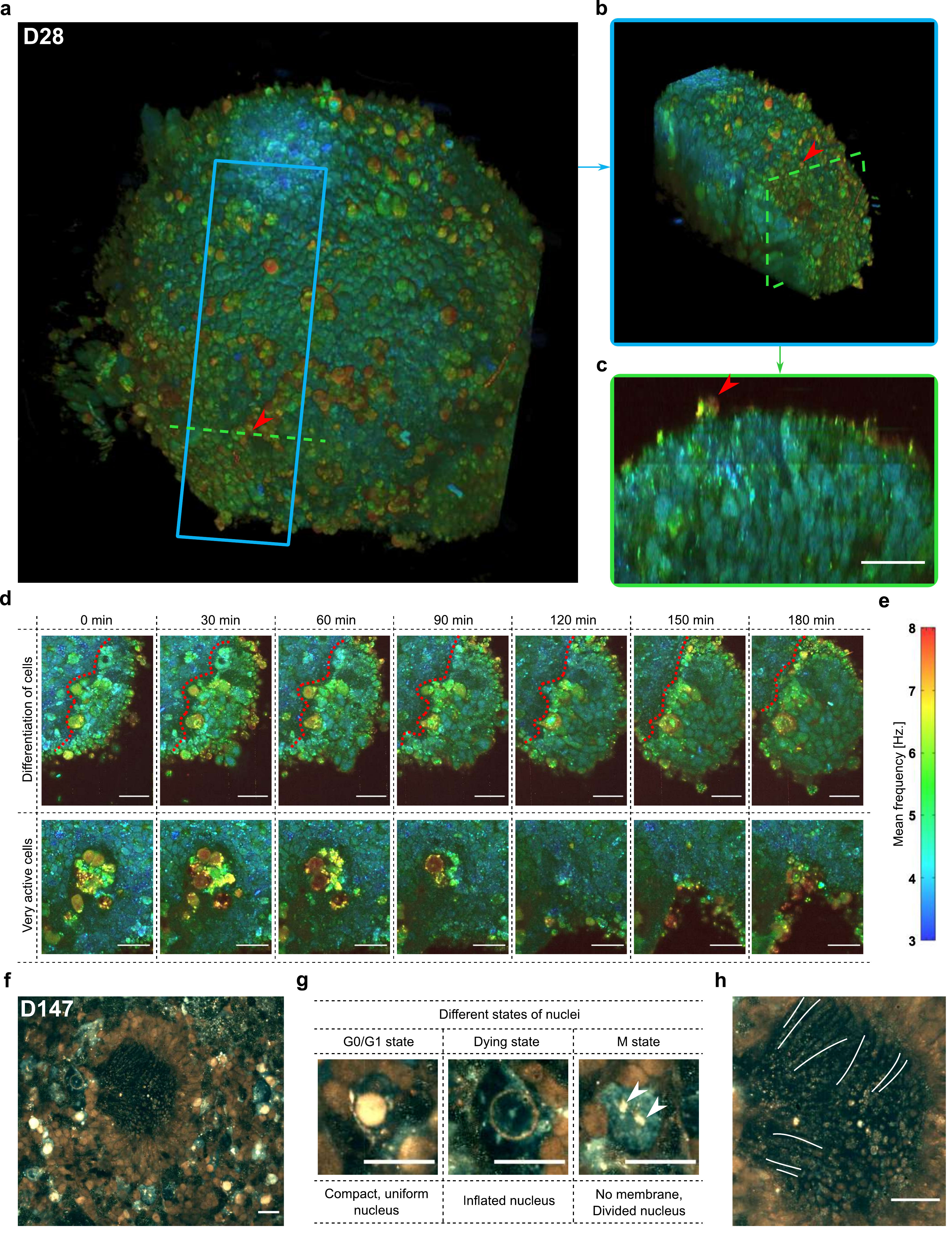}
    \caption{Imaging hiPSC-derived retinal organoids with D-FFOCT. \textbf{a}, 3D reconstruction of the spherical D28 retinal organoid, constituted of approximately $5~\mu m$ diameter cells. Red arrows highlight surface cells exhibiting fast dynamics. \textbf{b}, represents a sub-volume of \textbf{a} (blue square) and \textbf{c}, represents a cross-section in \textbf{a} (green dashed line) where one can see the organization of the layers inside the retinal organoid. \textbf{d}, Zooms on two different areas of the organoid during a 3h time-lapse acquisition: the top row zoom shows the differentiation of retinal progenitors into RPE (the boundary between the two types of cells is represented by a red dotted line), the bottom row shows a very active zone, composed of cells exhibiting fast and high dynamics, possibly undergoing apoptosis, in the center of the organoid. \textbf{e}, colorbar of the D-FFOCT images for the 3D and time-lapse acquisitions with a consistent colormap for \textbf{(a-d)}. High temporal resolution imaging performed on a D147 retinal organoid. \textbf{f}, Part of the retinal organoid revealing fusiform structures corresponding to emerging photoreceptor outer segments in the center of the rosette . \textbf{g}, zoom on nuclei in three different states around the rosette: (i) a nucleus in a normal state with a compact and uniform shape, very bright (i.e. exhibiting a high activity). (ii) what appears to be a dying inflated nucleus, with almost no activity in it and (iii) a nucleus in division with no defined nuclear membrane in the cytoplasm, and two distinct parts (white arrows) of the content of a nucleus (seems to be mitosis of the nucleus with chromosomes already divided, with the same subcellular activity level as the “normal” nucleus). \textbf{h}, zoom on the photoreceptor outer segment-like structures imaged side-on; three of them marked with a white line. Scale-bar: $20~\mu m$.}
    \label{fig2}
\end{figure}

A 3D reconstruction of a 28-day-old (D28) retinal organoid is depicted in Fig.~\ref{fig2}(a), corresponding to an optic vesicle stage during retinogenesis, along with a sub-volume in Fig.~\ref{fig2}(b) highlighting the layered internal retinal progenitor cell organization. A cross-section is shown in Fig.~\ref{fig2}(c) where the elongated shape of cells is seen. A time-lapse video at $50~\mu m$ depth was acquired on the same organoid in order to study its temporal evolution over three hours, see Fig.~\ref{fig2}(d) and Supplementary vid. 2 for the full recording. In these acquisitions, different dynamic behaviors of cells can be observed. Surface cells exhibit faster dynamics than those inside the sample volume. This could be explained by the fact that at the surface of the organoid, the cells are in contact with the external environment, making them more vulnerable to change and often leading to their death. In Fig.~\ref{fig2}(d) cells in the center of the organoid exhibit a fast and intense activity until their disappearance, possibly indicating that they are undergoing apoptosis. Evolution of cell dynamics near a clear boundary between two distinct types of cells is also visible. On one side of the boundary, cells differentiate into Retinal Pigment Epithelium (RPE) \cite{Reichman_pnas_2014}, exhibiting faster and stronger dynamics, and on the other side, small rounded progenitor cells have a slower activity. These two cell types are therefore distinguishable by their dynamic signature alone. Generated D-FFOCT images present a consistent colormap where each frequency is continuously represented by the same colors; therefore similar results are obtained for different retinal organoids at the same developmental stage. Supplementary Vid. 3 shows a timelapse movie of the D28 retinal organoids shown in Fig.~\ref{fig2}(a-c) alongside a D29 retinal organoids. The same clear boundary between distinct types of cells is present for both. By processing the data on the fly on the GPU using a modified version of Holovibes software \cite{holovibes} an enhanced temporal resolution of 50 ms was achieved, which represents a 500 fold improvement without the need of storing the raw data (up to $4~Go.s^{-1}$). The price paid for this improvement is the use of an alternative version of the dynamic computation which is noise-sensitive and non-quantitative (see Methods). Fig.~\ref{fig2}(f-h) shows high temporal resolution (20ms) images of a D7 retinal organoid. A typical rosette organization of retinal cells is visible in the center; i.e. photoreceptors (seen from the side) in the rosette center (indicated by white lines on Fig.~\ref{fig2}(h)), and other retinal cells surrounding. Photoreceptor nuclei exhibit different behaviors (Fig.~\ref{fig2}(g)): either compact and uniform; inflated; or absent, which may correspond respectively to the nuclear G0/G1, dying and M states. The gain in temporal resolution allows the study of fast biological processes such as organelles moving inside the cytoplasm, see Supplementary vid. 4. A series of retinal organoids imaged by D-FFOCT at consecutive steps of development showed the gradual differentiation of retinal cell progenitors into neural cells and photoreceptors Supplementary Fig.~\ref{fig1}, as validated by comparison with a similar organoid series imaged with immunofluorescence on a confocal microscope \cite{reichman_generation_2017}.

\begin{figure}
	\centering
    \includegraphics[width=1\linewidth]{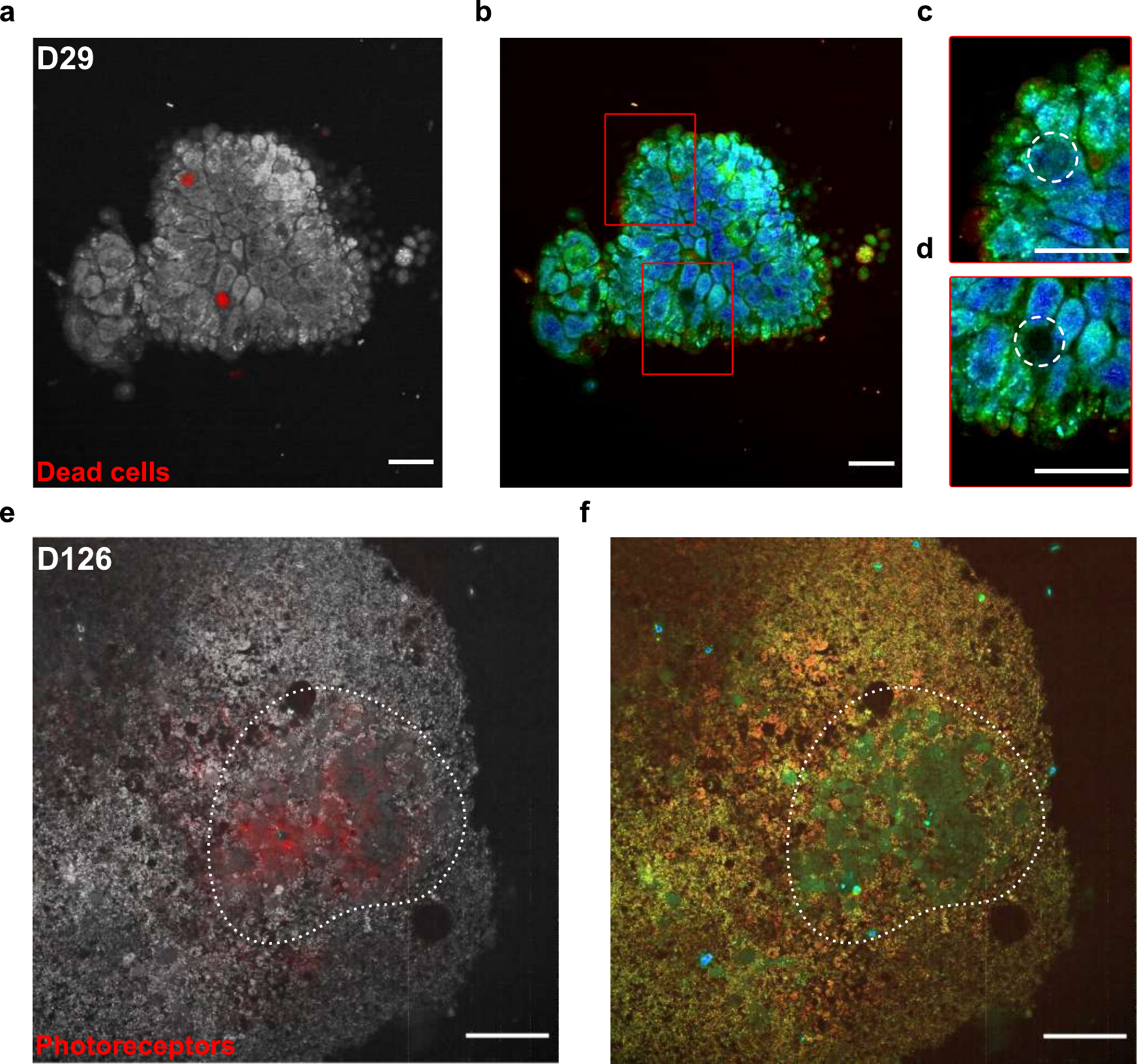}
    \caption{Fluorescence validation of D-FFOCT imaging. D-FFOCT images are depicted in color \textbf{(b-d, f)}, D-FFOCT images overlaid with wide-field fluorescence images are depicted in grayscale (D-FFOCT) and red (fluorescence) \textbf{(a, e)}. \textbf{(a-d)}, D29 retinal organoid labelled with a dye targeting the nuclei of dead cells.  \textbf{a}, One can see two dead cells marked by the red spots, corresponding to two dark zones on \textbf{b}, the D-FFOCT image, where there is no dynamic signal in these zones. \textbf{(c, d)}, show zooms on the two dark zones (highlighted by a white dashed line, corresponding to the two red spots of fluorescence). \textbf{(e-f)}, D126 retinal organoid derived from a fluorescent cone rod homebox (Crx) reporter iPSC line exclusively labelling photoreceptors in red (mCherry). \textbf{e}, Overlaid image on which the photoreceptor fluorescence matches the green cells of \textbf{f}, the D-FFOCT image. These areas are highlighted by a white dotted line. These precursors of photoreceptors have their own particular dynamic signature, which allows them to be distinguished from the surrounding cells by D-FFOCT alone. Scale bar: \textbf{(a-d)} $10~\mu m$ \textbf{(e, f)} $50~\mu m$.}
    \label{fig3}
\end{figure}

In order to further validate D-FFOCT signal origin via direct comparison between D-FFOCT and specific fluorescence labelling in the same organoids, a multimodal setup was developed which combines D-FFOCT and fluorescence channels to allow pixel-to-pixel overlay of D-FFOCT and fluorescence images. A 29-day-old retinal organoid was labelled with a dye targeting the nuclei of dead cells (see Methods). D-FFOCT images overlaid with fluorescence wide-field images are shown Fig.~\ref{fig3}(a-d). Two fluorescent red spots are clearly visible (Fig.~\ref{fig3}(a)) and correspond to very weak dynamic signals in the D-FFOCT image, confirming that dead cells exhibit low activity, and that the contrast revealed by D-FFOCT is metabolic. These two areas are zoomed in Fig.~\ref{fig3}(c,d) where the dark zones are circled by a white dotted line. To validate the identification of a specific cell population, we used retinal organoids derived from a photoreceptor-specific reporter iPSC line, in which nuclei of photoreceptor are intrinsically labelled in red with fluorescent protein mCherry \cite{gagliardi_characterization_2018}. A retinal organoid was imaged with the combined D-FFOCT-fluorescence system at D126, when a large number of differentiating photoreceptors can be detected in rosette-like structures (Fig.~\ref{fig3}(e, f)). A red fluorescent zone corresponding to photoreceptors is visible in Fig.~\ref{fig3}(e), whereas in the D-FFOCT image (Fig.~\ref{fig3}(f)), the different activity level in photoreceptors compared to surrounding cells is sufficient to provide distinction of the cell type through the dynamic signal alone, across a region that is coincident with the fluorescently labelled zone.

Dynamic FFOCT imaging creates a new label-free non-invasive contrast for imaging retinal organoids. As this technique does not damage the samples, it complements and could potentially replace the imaging modalities traditionally used. D-FFOCT allows imaging of different layers, at multiple depths, while preserving the sample integrity, i.e. using neither exogenous labelling nor destructive methods, and is therefore suitable to follow the evolution of the same organoid at different stages of its development. The high dimensionality of the probed signals (512 interferograms per pixel) is useful for developing statistical approaches such as automated classification and clustering, the only missing part for now being the lack of ground truth validation data (e.g. segmented cells with labels that could be generated by fluorescence or by annotating experts) which will be a milestone in the further development of this technique.

\section*{Acknowledgements}

This work was supported by grants entitled “HELMHOLTZ” (European Research Council (ERC) (\#610110), PI Mathias Fink and José-Alain Sahel), “LABEX LIFESENSES” [ANR-10-LABX-65] supported by the ANR within the Investissements d'Avenir program [ANR-11-IDEX-0004-02],  (PI Olivier Goureau), and “OREO” (ANR),  (PI Kate Grieve). The authors would like to thanks Amélie Slembrouck-Brec and Céline Nanteau for their contribution in the preparation of the samples, Olivier Thouvenin and Pedro Mecê for fruitful discussions.

\section*{Author contributions}

J.S. wrote the software for experiments and data processing. J.S. and K.Gro. performed the experiments and analysed the results. J. S., K.Gro., C.B. and K.Gri. devised and designed the experiments driven by inputs from S.R. and O.G.. O.G. and S.R. provided the samples. All authors discussed the results and contributed to the manuscript.

\section*{Competing interests}

The authors declare that they have no competing interests in this study.

\newpage



\section*{Methods}

\subsection*{Human iPSC maintenance and retinal differentiation}

Two established human iPSC lines, hiPSC line-5f \cite{slembrouck-brec_reprogramming_2019} and fluorescent reporter \\ AAVS1::CrxP\_H2BmCherry hiPSC line \cite{gagliardi_characterization_2018} both derived from retinal Müller glial cells were cultured as previously described \cite{reichman_generation_2017}. Briefly, hiPSC lines were cultured on truncated recombinant human vitronectin-coated dishes with Essential 8TM medium (ThermoFisher Scientific). For retinal differentiation, adherent hiPSCs were expanded to 70-80 \% and FGF-free medium was added to the cultures for 2 days, followed by a neural induction period allowing the appearance of retinal structures. Identified retinal organoids were manually isolated and cultured as floating structures for several weeks to follow retinal differentiation as previously described \cite{Reichman_pnas_2014, reichman_generation_2017, slembrouck-brec_defined_2018}.

\subsection*{Sample preparation}

Retinal organoids were placed in CO2 independent medium (GibcoTM, Fisher Scientific) in a petri-dish and kept close to $37^\circ C$ during imaging. Samples were mounted on a 3-axis translation stage under the sample arm objective and imaged directly after mounting. After imaging, each organoid was either cultured again for further D-FFOCT imaging or fixed using a solution of paraformaldehyde (PFA) for 10 minutes at $4^\circ C$ followed by three rinces and stored in a sucrose solution for further traditional imaging. Dead cell labelling (LIVE/DEAD Viability/Cytotoxicity kit for mammalian cells, Invitrogen molecular probes life technologies) with $10~ \mu M$ EthD-1 was incubated at $37^\circ C$ for 20 minutes before imaging. Human iPSC samples were mounted immediately after incubation and imaged within 10 minutes.

\subsection*{Immunostaining and imaging on retinal sections}

For cryosectioning, retinal organoids were fixed for 15 min in 4$\%$ paraformaldehyde (PAF) at 4$\degree$C and washed in Phosphate-buffered saline (PBS). Structures were incubated at 4$\degree$C in PBS / 30$\%$ Sucrose (Sigma-Aldrich) solution during at least 2 hours and embedded in a solution of PBS, 7.5$\%$ gelatin (Sigma-Aldrich), 10$\%$ sucrose and frozen in isopentane at -50$\degree$C. 10~$\mu m$-thick cryosections were collected in two perpendicular planes. Sections were washed with PBS, nonspecific binding sites were blocked for 1 hour at room temperature with a PBS solution containing 0.2$\%$ gelatin and 0.1$\%$ Triton X-100 (blocking buffer) and then overnight at 4$\degree$C with the primary antibody VSX2 (Goat, 1:2000, Santa Cruz), CRX (mouse, 1:5000, Abnova) and RHODOPSIN (Mouse, 1:500, Millipore) diluted in blocking buffer. Slides were washed three times in PBS with 0.1$\%$ Tween and then incubated for 1 hour at room temperature with appropriate secondary antibodies conjugated with either Alexa Fluor 488 or 594 (Interchim) diluted at 1:600 in blocking buffer with 4',6-diamidino-2-phenylindole (DAPI) diluted at 1:1000 to counterstain nuclei. Fluorescent staining signals were captured with an Olympus FV1000 confocal microscope.

\subsection*{Experimental setup}

FFOCT \cite{beaurepaire_98, dubois_ultrahigh-resolution_2004} is a variant of regular OCT \cite{huang_optical_1991} in which two-dimensional en face images are captured on a CMOS camera. Three-dimensional images can be acquired and reconstructed by scanning in the depth direction with high precision motors. This configuration, together with the use of a broad-band LED source, allows for higher axial and en face resolution than conventional OCT. Moreover, FFOCT can perform micrometer resolution 3D imaging non-invasively in both fresh or fixed ex vivo tissue samples. For the D-FFOCT imaging of retinal organoids, a laboratory setup was designed with a $0.5~\mu m$ lateral resolution, using high-magnification water-immersion objectives (Nikon NIR APO 40x 0.8 NA), for a field of view of approximately $320\times320~ \mu m^2$. Because of the high numerical aperture, the axial resolution of $1.7~\mu m$ is also given by the microscope objectives in this particular configuration where the coherence length of the source is larger than the depth of focus. The source used for the FFOCT system was an LED centered at $660~nm$ (M660L3, Thorlabs, Newton, NJ, USA). The FFOCT signal is recorded on a custom camera (Quartz 2A750, Adimec). For validation purposes, this FFOCT setup was combined with a fluorescence microscope, using an LED source centered at $565~nm$ (M565L3, Thorlabs, Newton, NJ, USA) for the excitation and filtered with a band-pass filter centered on $562~nm$ with a bandwidth of $40~nm$ (Semrock FF01-562/40-25). The emitted fluorescence is filtered with another band-pass filter centered on $624~nm$ (Semrock FF01-624/40-25) and then imaged on a sCMOS camera (PCO Edge 5.5). The excitation and fluorescence wavelengths are separated by a dichroic mirror at $593~nm$ (Semrock FF593-Di03-25).

\subsection*{Data acquisition and processings}

Producing each dynamic image slice requires acquisition of many (typically 512) direct images without modulating the piezo position. As opposed to static FFOCT acquisition, the measured fluctuations in D-FFOCT arise from sub-cellular motion. In this study the frame rate was set to $100~Hz$ which was a good trade-off between acquisition speed and signal to noise ratio. For a given acquisition we typically obtain a $(1440, 1440, 512)$ tensor where $1440 \times 1440$ is the sensor number of pixels and 512 is the number of recorded frames. After acquiring the data, the first step is to correct for the camera frame-to-frame instability by normalizing each frame, in order to compensate for exposure time variations. Previously, we showed color images that were constructed by integrating signals in the Fourier domain for three frequency ranges \cite{apelian_extracting_2017}. Here we propose a new scheme where color images are computed in the Hue-Saturation-Value (HSV) colorspace where, contrary to the Red-Green-Blue color space, it is possible to assign a physical property to each of the three channels. The idea is then to attribute a color for each pixel depending on the characteristic time period or frequency of the dynamic signal. Each individual pixel can be thought of as a sum of subcellular random walk with a typical covariance function depending on the motion type (e.g. diffusive, hyper-diffusive). To distinguish between several behaviors, we perform a power spectrum analysis. Note that a correlation analysis could also be made and gives similar results but requires more computing time. We start by computing the power spectrum density (PSD) using Welch’s method for each pixel and then L1 normalize each PSD as if it were a probability distribution. Then the Hue channel is computed as the mean frequency and is simply the dot product between the normalized power spectrum density and the frequency array. The values are then inverted and rescaled between 0 and 0.66 in order to go from blue (low frequencies) to red (high frequencies). We observed that two successive acquisitions could lead to different perceptual colormaps. We found that the first frequencies were slightly unstable (either due to sensor or mechanical instabilities as described in [Scholler, Optics Express, 2019]). We removed this artifact by removing the first 3\% of the frequencies in the PSD. Then, the value which corresponds to the perceived pixel intensity is computed as the average of the temporal standard deviation with a window size of 50 samples. We saturate 0.1\% of the highest value pixels to improve the contrast. For 3D stacks the saturation value is kept the same throughout the stack to obtain a consistent colormap. Finally, the saturation is computed as the inverse of the normalized power spectrum density bandwidth. As a consequence, the saturation channel carries the frequency bandwidth information and is simply computed as:
\begin{align}
   S =  \boldsymbol{P}.\boldsymbol{f}^2-(\boldsymbol{P}.\boldsymbol{f})^2
\end{align}
Where P is the normalized power spectrum density array, f is the frequency array and "." is the dot product. The saturation map is then inverted and rescaled between 0 and 0.8 in order to obtain a visually pleasing output. The broader the spectrum, the lower the saturation. White noise has the broader bandwidth and will therefore appear grayish instead of colored. Finally the dynamic image is transformed in the RGB colorspace for display purposes. A (1440; 1440; 512) stack is processed in less than 10 seconds by the GPU (Nvidia Titan V) using Matlab (MathWorks) hence limiting the temporal resolution by processing the data after each acquisition. In order to improve the temporal resolution, we also used a modified version of \cite{holovibes} which computes the dynamic images on the GPU in real-time. In this case, the algorithm used to generate dynamic images is no longer quantitative and is subject to more noise. The 3D cube of data is first Fourier transformed, the Fourier transform of each temporal signal is then integrated into three bands in order to reconstruct a Red Green Blue (RGB) image where the band corresponding to lower frequencies is coded in the Blue channel, the band corresponding to medium frequencies is coded in the Green channel and the band corresponding to high frequencies is coded in the Red channel. For 3D stacks, each plane is registered in post-processing using a feature based method (rigid registration using sift features and RANdom SAmple Consensus (RANSAC) algorithms to find correct matches) in order to compensate for the sample lateral drift. Stacks are then interpolated in the depth direction using bicubic interpolation in order to obtain a square voxel edge size of $220~nm$. For 3D display purposes, a non-local mean filter \cite{buades_non-local_2005} can be applied to remove granularity.

\subsection*{Correcting for sample drift}

Due to thermal effects, we observed a slow mechanical drift (on the order of $5~mu m.h^{-1}$) which could prevent us from measuring dynamic images in the same plane over several hours (axial resolution is $1.7~mu m$). In order to compensate for this drift, we developed a correlation based method which we called the plane locking procedure. This procedure was triggered when the cross-correlation between the current image and the target image was below a threshold (typically between 0.2-0.4). In this case, FFOCT images are acquired over an axial extent of $10~\mu m$, with $0.5~\mu m$ steps using the sample translation stage, and are then cross-correlated with the target image. The sample is then axially translated to the position corresponding to the maximum of the cross-correlation, which was typically between 0.5-0.8. After each plane correction procedure, a new FFOCT image is taken as target for the next correction in order to account for evolution in the sample position.

\subsection*{Combining FFOCT and fluorescence microscope channels}

The FFOCT and fluorescence cameras do not share the same sensor size and resolution so in order to construct overlays, FFOCT images were registered onto fluorescence images using a projective transformation. The projective transformation was calibrated using a deformation target (R1DS1N, Thorlabs, Newton, NJ, USA) before experiments. The final overlays are constructed in the RGB colorspace, the Value channel corresponding to dynamic amplitude was put into every channel (R, G and B) and the fluorescence image was added only in the R channel. In this way, the dynamic image appears in grayscale with the fluorescence superimposed in red.

\subsection*{Data availability}
The study data are available from the corresponding author upon request.

\subsection*{Code availability}
The control and acquisition software is available on \url{https://github.com/JulesScholler/FFOCT} \cite{scholler_zenodo_2019}.



\newpage

\bibliographystyle{ieeetr}
\bibliography{biblio_organoids}

\newpage



\section*{Integrated supplementary information}

\begin{figure}[H]
    \centering
    \includegraphics[width=1\linewidth]{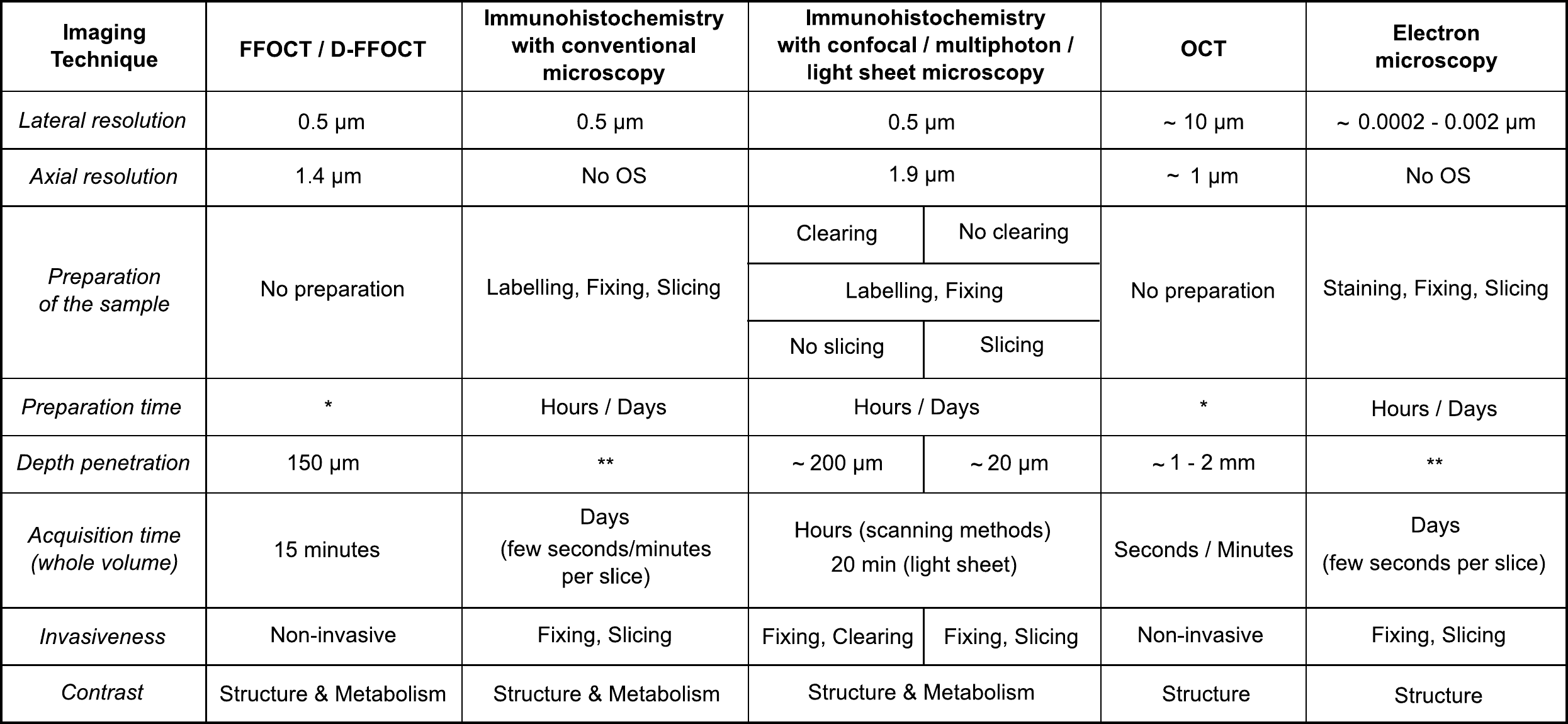}
    \captionsetup{format=sanslabel}
    \caption{Table 1: Comparison of the different techniques which can be used for the imaging of hiPSC-derived retinal organoids. OS: optical sectioning. *: As there is no optical sectioning for these techniques, the slices must be thin i.e. on the order of the depth of field ($0.5-3~\mu m$). When the slice is considered "thick", it measures from ~$10$ to ~$100~\mu m$. 1: The resolutions have been calculated for the same microscope objectives as those used in this study (Nikon NIR APO 40x 0.8 NA). Current imaging modalities for iPSC-derived retinal organoids include immunohistochemistry methods, optical coherence tomography (OCT) and electron microscopy. Immunohistochemistry methods rely either on staining which adds specific color contrast and where the sample is mechanically sliced to provide sectioning (immunohistology) or with targeted dyes where various techniques (confocal illumination/detection, light-sheet illumination, etc.) provide optical sectioning (immunofluorescence). Relying on exogenous contrasts for imaging greatly limits what can be seen in samples and ultimately leads to the sample’s destruction, as these techniques all require slicing, clearing or fixation steps which preclude the study of the same sample during its development. An advantage of immunochemistry however is that it reliably provides interpretable images with a diffraction limited resolution. Traditional OCT suffers from a trade-off between the Depth Of Field (DOF) that is needed for imaging 3D samples and the lateral resolution. In order to have ~$100~mu m$ DOF the lateral resolution is limited to typically $10~\mu m$ which is not sufficient to resolve cellular structures. The benefit of using a full field configuration is that the numerical aperture can be arbitrarily high without compromising the depth imaging capabilities and can therefore benefit from the latest advances in microscope objective technology. Electron microscopy offers very high resolution which overcomes the diffraction limit due to the near field nature of the measurement. The field of view however is very small and not adequate for samples with thousands of cells. The characteristics of each of the described techniques used for imaging of retina organoid are shown Table 1. D-FFOCT is the only technique that combines high spatial and temporal resolution, which already make it interesting, but it is also non-destructive which opens up the possibility to follow the development of the same organoids throughout development or drug screening, a process which could be automated by integrating the D-FFOCT setup directly inside the incubator.}
    \label{table1}
\end{figure}

\begin{figure}[H]
    \centering
    \includegraphics[width=1\linewidth]{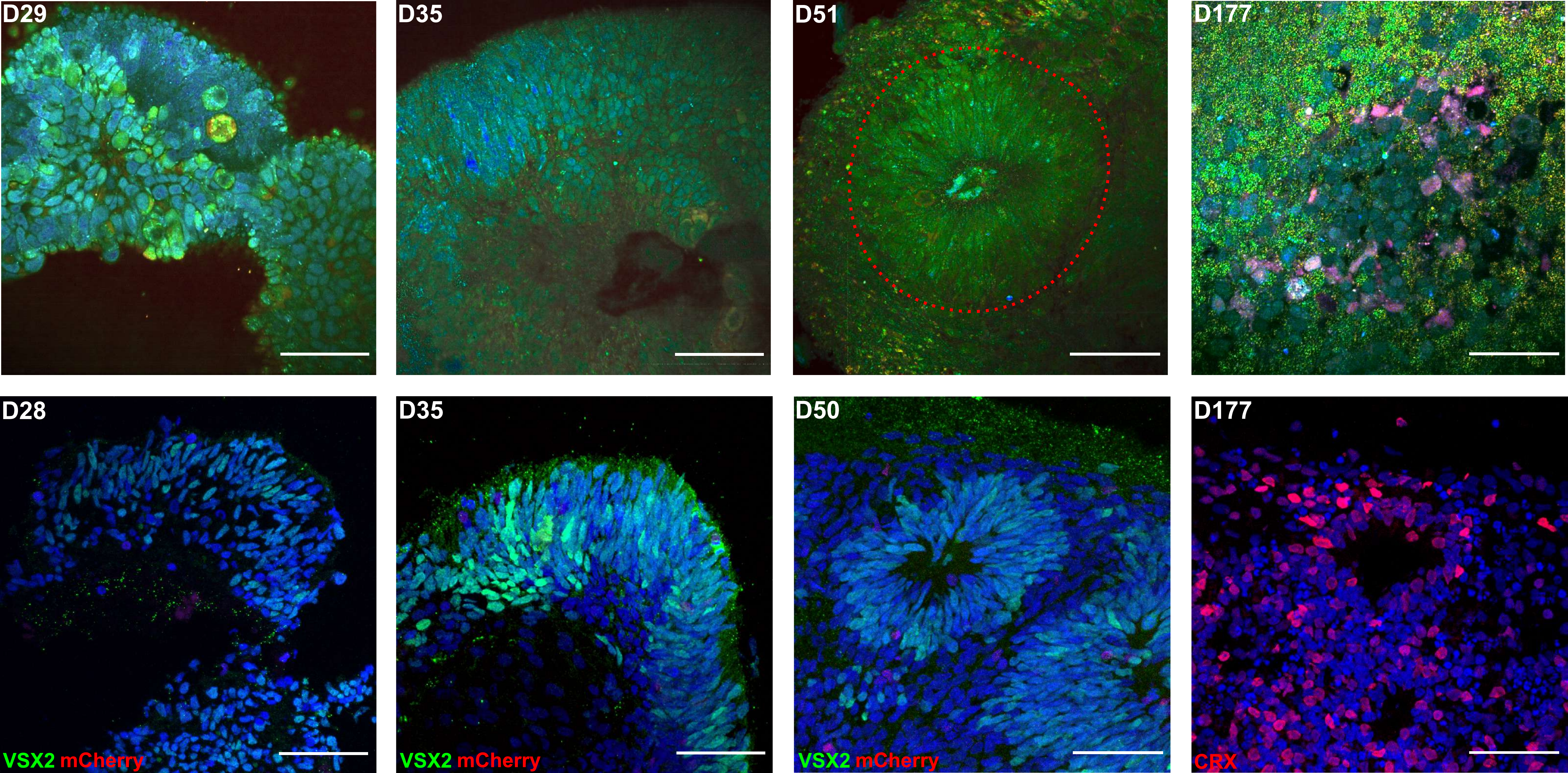} 
    \captionsetup{format=sanslabel}
    \caption{Figure 1: D-FFOCT imaging of four different retinal organoids at different stages of development follows organoid differentiation (appearance of photoreceptors). At D29 and D35, organoids are mainly composed of multipotent retinal progenitors cells expressing Visual System Homeobox 2 (VSX2) transcription factor (green). At D51, rosettes (shown by red dotted line) containing photoreceptor precursors begin to appear. Finally, at D177, mature photoreceptors expressing CRX are present in rosette structures. Below each D-FFOCT image, the corresponding immunhistochemistry images (from different retinal organoids but at a similar age), acquired with confocal microscope, are presented. Nuclei are counterstained with DAPI (blue). The first three organoids (D28, D35, D50) are derived from the iPS 5FC organoid line, expressing mCherry on the CRX gene endogenously. At D177, CRX was shown by immunostaining on a non-fluorecent reporter organoid. (Here, the field of view of D-FFOCT images has been reduced to correspond to the field of view of the confocal images). Scale-bar: $50~\mu m$.}
    \label{fig:supp_1}
\end{figure}

\begin{figure}[H]
    \centering
    \includegraphics[width=1\linewidth]{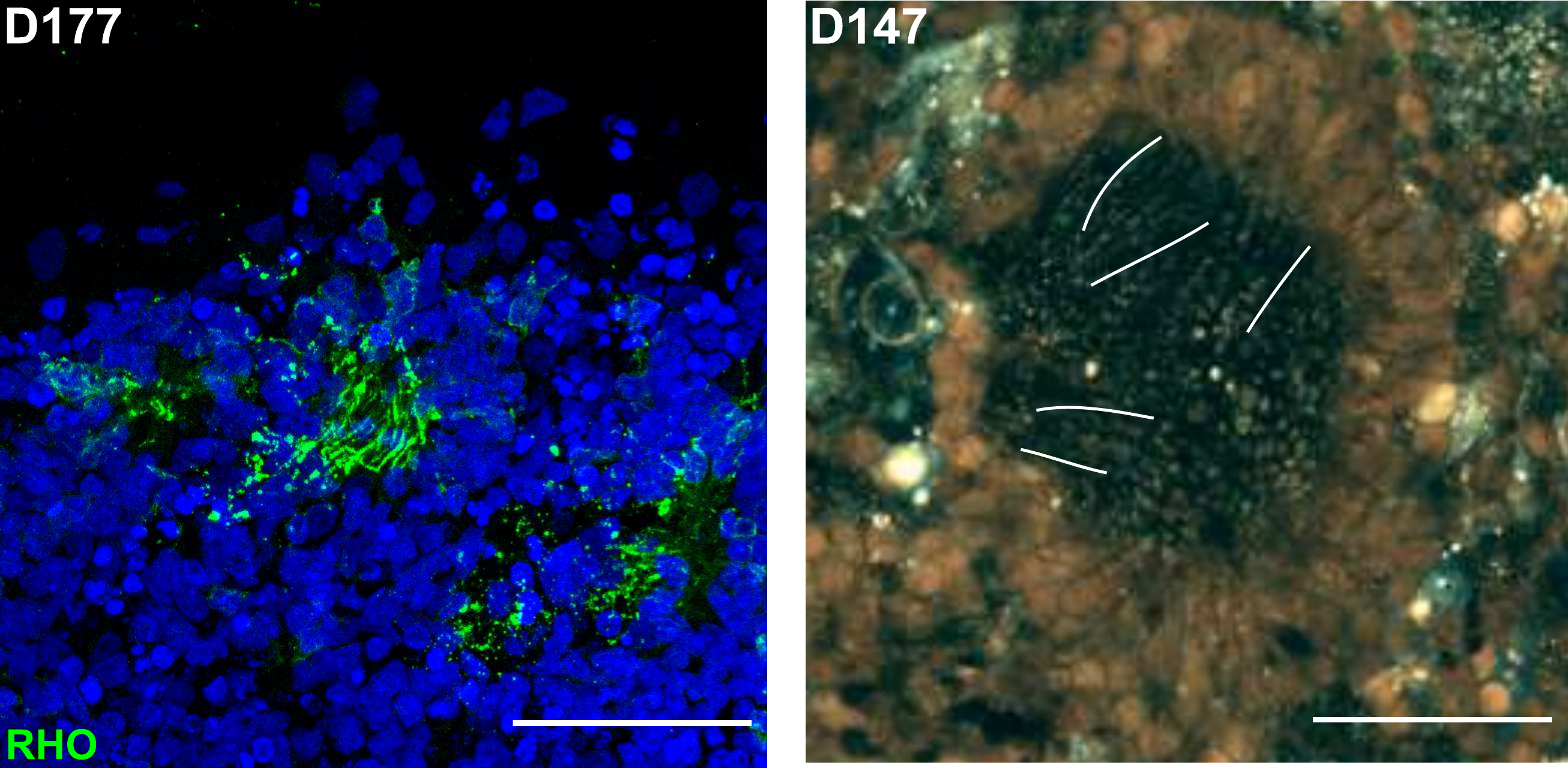} 
    \captionsetup{format=sanslabel}
    \caption{Figure 2: Immunohistochemistry image (left), acquired with confocal microscope, of a D177 retinal organoid, highlighting the outer segments of photoreceptors containing rhodopsin (green). nuclei are counterstained with DAPI (blue). D-FFOCT image of a 147-day-old retinal organoid (right) showing the center of a rosette which is composed of photoreceptors' outer segments (some of them are highlighted by a white line), showing the same shape as in the immunohistochemistry image. The field of view of the D-FFOCT image has been reduced to coincide with the scale of the immunohistochemistry image.  Scale-bar: $50~\mu m$.}
    \label{fig:supp_2}   
\end{figure}

\newpage 
\section*{Supplementary information}

\paragraph{Video 1} Depth stack of the D28 hiPSC-derived retinal organoid shown in Fig. 2(a-c). Planes from top ($0~\mu$m) to approximate center of the organoid ($101~\mu$m), in $1~\mu$m steps are shown revealing the internal structure of the organoid. 

\paragraph{Video 2} Time-lapse video with 1 min temporal resolution of the evolution of the D28 hiPSC-derived retinal organoid shown in Fig. 2(d). Video shows the evolution of the whole organoid over three hours of imaging, with three zooms corresponding to the zones depicted in Fig. 2(d). 

\paragraph{Video 3} Time-lapse video with 1 min temporal resolution of the evolution of the D29 hiPSC-derived retinal organoid shown in Supplementary Fig. 2. Video shows the evolution of the whole organoid over three hours of imaging. 

\paragraph{Video 4} Time-lapse of the D29 hiPSC-derived retinal organoid and of the D28 hiPSC-derived retinal organoid. Both organoids exhibit the same consistent colormap and structures demonstrating that D-FFOCT can reliably images different samples at different times consistently. 

\paragraph{Video 5} Time-lapse video with 1 min temporal resolution of the evolution of a D42 hiPSC-derived retinal organoid. Video shows the evolution of the whole organoid over three hours of imaging, especially the behaviour in a rosette. 

\paragraph{Video 6} Time-lapse video with $5~s$ temporal resolution of the evolution of a D147 hiPSC-derived retinal organoid with zoomed areas. Zoomed areas have a temporal resolution of 1 second and highlight the sensitivity of D-FFOCT to dynamic phenomena. The colors are calculated like the images of Fig. 2(f-h).


\end{document}